\documentclass{aastex61}
\received{May 17, 2017}
\revised{May 23, 2017}
\accepted{May 25, 2017}
\submitjournal{ApJL}

\shorttitle{Increasing amplitude of small-scale LOS velocities before a filament eruption}
\shortauthors{Seki et al.}
\begin{document}
\title{INCREASE IN THE AMPLITUDE OF LINE-OF-SIGHT VELOCITIES OF THE SMALL-SCALE MOTIONS IN A SOLAR FILAMENT BEFORE ERUPTION}

\correspondingauthor{Daikichi SEKI}
\email{seki@kwasan.kyoto-u.ac.jp}

\author{Daikichi SEKI}
\affil{Graduate School of Advanced Integrated Studies in Human Survivability, Kyoto University, Sakyo, Kyoto 606-8306, Japan}
\affiliation{Kwasan and Hida Observatories, Kyoto University, Yamashina, Kyoto 607-8471, Japan}

\author{Kenichi OTSUJI}
\affiliation{Kwasan and Hida Observatories, Kyoto University, Yamashina, Kyoto 607-8471, Japan}

\author{Hiroaki ISOBE}
\affil{Graduate School of Advanced Integrated Studies in Human Survivability, Kyoto University, Sakyo, Kyoto 606-8306, Japan}

\author{Takako T. ISHII}
%\altaffiliation{Creator of AASTeX v6.1}
\affiliation{Kwasan and Hida Observatories, Kyoto University, Yamashina, Kyoto 607-8471, Japan}
%\collaboration{(LaTeX collaboration)}

\author{Takahito SAKAUE}
%\altaffiliation{Creator of AASTeX v6.1}
\affiliation{Kwasan and Hida Observatories, Kyoto University, Yamashina, Kyoto 607-8471, Japan}
%\collaboration{(LaTeX collaboration)}

\author{Kumi HIROSE}
\affiliation{Kwasan and Hida Observatories, Kyoto University, Yamashina, Kyoto 607-8471, Japan}



\begin{abstract}

We present a study on the evolution of the small scale velocity field in a solar filament as it approaches to the eruption. 
The observation was carried out by the Solar Dynamics Doppler Imager (SDDI) that was newly installed 
on the Solar Magnetic Activity Research Telescope (SMART) at Hida Observatory. 
The SDDI obtains a narrow-band full disk image of the sun at 73 channels from H$\alpha$ - 9.0 \AA\ to H$\alpha$ + 9.0 \AA, 
allowing us to study the line-of-sight (LOS) velocity of the filament before and during the eruption. 
The observed filament is a quiescent filament that erupted on 2016 November 5. 
We derived the LOS velocity at each pixel in the filament using the Becker's cloud model, 
and made the histograms of the LOS velocity at each time. The standard deviation 
of the LOS velocity distribution can be regarded as a measure for the amplitude 
of the small scale motion in the filament. 
We found that the standard deviation on the previous day of the eruption was 
mostly constant around 2--3 km s$^{-1}$, and it slightly increased to 3--4 km s$^{-1}$ 
on the day of the eruption. It shows further increase with a rate of 1.1 m s $^{-2}$
about three hours before eruption and again with a rate of 2.8 m s $^{-2}$ 
about an hour before eruption. 
From this result we suggest the increase in the amplitude of the small scale motions
in a filament can be regarded as a precursor of the eruption.

\end{abstract}




\section{Introduction} \label{sec:intro}

Dark filaments, or prominences, dense cooler plasmas supported by magnetic field in the solar corona, often become unstable and erupt. Filament eruption is associated with various phenomena such as flares, coronal mass ejections (CMEs) and giant arcade formations in the quiet sun. Despite their diversity in size, morphology, and emitting radiation spectrum, they are considered to be the different aspects of the common magnetohydrodynamical processes that involve plasma ejection and magnetic reconnection  \citep{2011LRSP....8....6S}.

Filament eruptions often follow the slow rise \citep{2004ApJ...602.1024S, 2011ApJ...743...63S}, plasma motions \citep{2006A&amp;A...449L..17I, 2009SoPh..259...13G}, weak heating \citep{2006A&amp;A...458..965C}, and increased internal motions \citep{tandberg1995nature} in the filament. 
In more general context of the solar eruptions, various kinds of "precursors" have been proposed, 
including emerging magnetic flux \citep{1995JGR...100.3355F, 2000ApJ...545..524C, 2012ApJ...760...31K} 
magnetic reconnection at various magnetic configuration \citep{1999ApJ...510..485A, 2001ApJ...552..833M}, 
and helicity injection \citep{2008PASJ...60.1181M, 2009ApJ...691L..99H}. 
Of particular interest for the present study is the pre-flare increase in the the non-thermal velocity.  
By the spectroscopic observation in the the coronal line, \citep{2001ApJ...549L.245H} found 
the non-thermal line broadening before the increase in the X-ray flux and the electron temperature,  
suggesting the increase in the turbulent motion taking place before the onset of the flare.  

In this Letter, we report the observation of a quiescent filament eruption by the Solar Dynamics Doppler Imager
(SDDI)  \citep{2017SoPh..292...63I} newly installed on the on the Solar Magnetic Activity Research Telescope (SMART) \citep{2004SPIE.5492..958U} at Hida Observatory, Kyoto University. The SDDI takes full disk solar images at 73 wavelength positions around H$\alpha$ which allows us to monitor the H$\alpha$ line profile of the filament prior to and during an eruption.

\section{Observation} \label{sec:obs}
The Solar Dynamics Doppler Imager (SDDI) installed on the Solar Magnetic Activity Research Telescope (SMART) at Hida Observatory of Kyoto University has been operating a routine observation since 2016 May 1.  It takes the solar full disk images of 73 channels at every 0.25 \AA from H$\alpha$ line center - 9.0\AA \ to H$\alpha$ line center + 9.0\AA, i.e. at 36 positions in the blue wing, H$\alpha$ line center, and 36 positions in the H$\alpha$ red wing. Each image is obtained with the time cadence of 15 seconds and the pixel size of about 1.2 arcsec \citep{2017SoPh..292...63I}. When the weather permits, it continuously monitor the sun during the day time in Japan for 10 hours. The detail of the instrument and the examples of images and the line profiles can be found in \cite{2017SoPh..292...63I}.

\begin{figure}[ht!]
\gridline{\fig{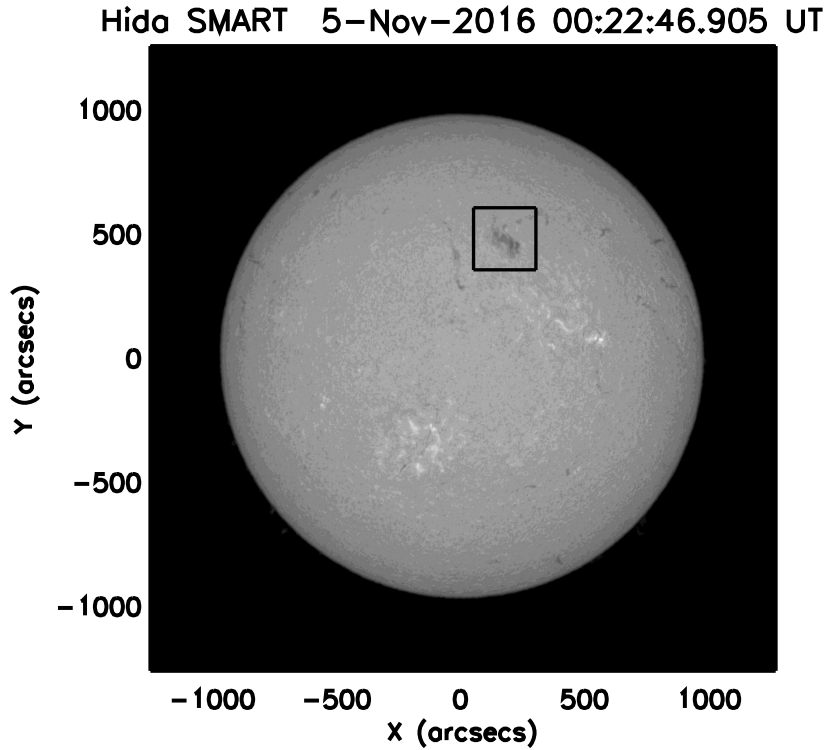}{0.6\textwidth}{}}
\caption{A solar full disk image in H$\alpha$ line center captured by Solar Dynamics Doppler Imager(SDDI) installed in Solar Magnetic Activity Research Telescope(SMART) at Hida Observatory, Kyoto University. The filament shown inside the black square is the target filament analyzed in this research.\label{fullsun}}
\end{figure}

In this Letter we used the SDDI images taken from 23:00UT on 2016 November 3 to 7:00UT on 2016 November 4 and from 22:00UT on 2016 November 4 to 5:00UT on 2016 November 5.  Figure\ref{fullsun} shows the full disk image at  H$\alpha$ center at 00:22:46UT on November 5. 

\begin{figure}[ht!]
\gridline{\fig{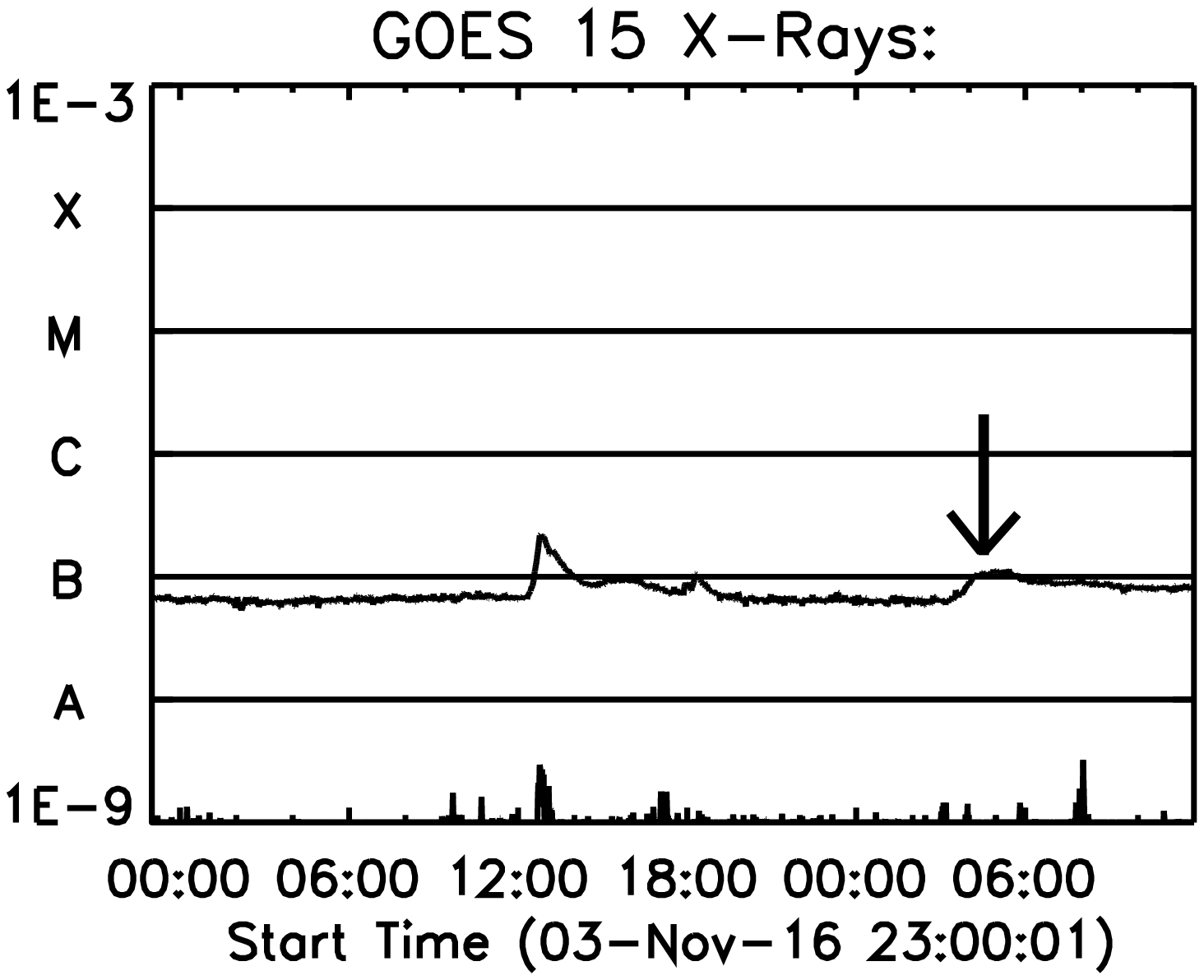}{0.7\textwidth}{}}
\caption{Soft X-ray flux is plotted from GOES15 from 23:00UT on 2016 November 3 to 12:00UT on 2016 November 5. The upper line shows the long X-ray flux (between 1.0 \AA and 8.0 \AA), while the lower lines shows the short X-ray flux (between 0.5 \AA and 4.0 \AA). The red arrow indicates the time when the filament erupted. \label{goes}}
\end{figure}

The filament was located in the quiet sun in the northern hemisphere, and it erupted around 3:30UT on November 5. There are weak active regions including NOAA 12605 on the sun, but the overall solar activity was very low. Thanks to the low activity, the eruption can be barely identified as a B1-class event in the GOES soft X-ray flux, shown in Figure \ref{goes}. 

The gradual enhancement starting around 4:00UT and peaking around 4:50UT corresponds to the filament eruption. In the EUV images from the SDO/AIA  \citep{2012SoPh..275...17L}, formation of a giant arcade after the filament eruption can be found (not shown in the figures). The erupted filament eventually became a slow CME, which seems to be the cause of the moderate disturbance of the geomagnetic field from November 9 to November 10. 

%We applied Beckers' cloud model \citep{1964beckers, 1988A&amp;A...203..162M} to the SDDI data to determine source function, doppler width, doppler shift, and optical depth of the filament \citep{2003PASJ...55..503M, 2003PASJ...55.1141M, 2010PASJ...62..939M, 2017ApJ...836...33C}. The wide wavelength coverage and the high spectral resolution of the SDDI allow us to determine the physical parameters precisely even during the filament activation or eruption when the line profile shifts from the nominal line center (6563 \AA) significantly. From the doppler shift, we obtain the line-of-sight (LOS) velocity at each pixel in the filament, and the LOS velocity images are shown in Figure \ref{5types}.

\begin{figure}[ht!]
% \centering
% \includegraphics[width=3cm,clip]{1104-1105_stddev.eps}
\plotone{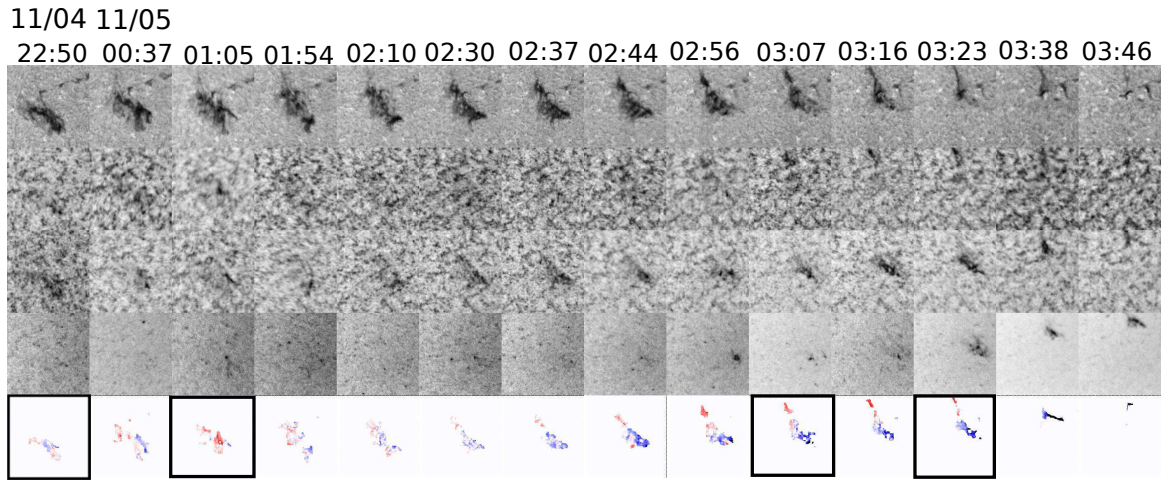}
\caption{Time series of H$\alpha$ images at the line center, at + 0.5 \AA, at - 0.5 \AA,  and at - 1.0 \AA (the top three row) and of the images of LOS velocity of the filament (bottom row). The velocity scale of the bottom images are identical to the right images of Figure \ref{4hist}. \label{5types}}
\end{figure}

We applied the Beckers' cloud model \citep{1964beckers, 1988A&amp;A...203..162M} to the SDDI data. By applying the model to the images taken at multiple channels around H$\alpha$ we can determine source function, doppler width, doppler shift, and optical depth of filaments. The plane-of-sky motion in the images and the doppler shift derived from the model can be used to calculate the three-dimensional velocity of erupting filaments \citep{2003PASJ...55..503M, 2003PASJ...55.1141M, 2010PASJ...62..939M, 2017ApJ...836...33C}. 
The wide wavelength coverage and the high spectral resolution of the SDDI allow us to determine the physical parameters precisely even when the line profile shifts from the nominal line center (6563 \AA) significantly. 
From the doppler shift, we obtain the line-of-sight (LOS) velocity at each pixel in the filament, and the LOS velocity images are shown in Figure \ref{5types}.

To derive the LOS velocity at each pixel, we conducted three automatic steps to determine the binary images which can cover most of the pixels inside the filament (we call them ``masks''). First, we determined the positions of the pixels where the intensities are lower than $I_m - 2 \sigma_I$ ($I_m$ is the average of the intensities inside the black square area shown in Fig. \ref{fullsun} and $\sigma_i$ is their standard deviation) for each wavelength image and got all the positions together to obtain the binary image whose pixels at the same positions have 1 and the other pixels have 0. After that, we smoothed the obtained images by taking the average of 5x5 pixels around each pixel and select the pixels whose values are 1. The last step is to operate ``dilation'' and ``erosion'' processes. Dilation is a process that if at least one of the surrounding pixels is 1 for a certain pixel, the pixel will become 1, i.e. it will be 0 only if all the 8 pixels  around it are 0. On the other hand, erosion is a opposite process that a certain pixel will be 1 only if all the surrounding 8 pixels are 1. By utilizing this fundamental morphological image processing, we operated 4 processes, erosion-dilation-dilation-erosion in order, to the images after the second step. The last 2 steps (smoothing process and erosion-dilation-dilation-erosion process) are necessary to remove the spot noises appearing in the images after 1st step produced by the spicules in the images of $\pm$ 0.5 \AA. 

These 3 steps gave us masks that covers most of the filament, and we operated the cloud model fitting to the pixels inside the mask for each image. In this study we will focus on the LOS velocity and not use the other parameters derived from the model. 

Figure \ref{5types} shows the time series of the SDDI images. From the top to the bottom: H$\alpha$ center, +0.5 \AA\,, -0.5  \AA\,, -1.0 \AA\,, and the LOS velocity. 

The field of view is indicated as the black square in Figure \ref{fullsun}. The filament was stable and only the small portions were barely visible at $\pm$ 0.5 \AA, before 00:00UT on Nov. 5. However, from $\sim$ 00:30UT the small scale motions in the filament became noticeable in the wing images as well as the LOS velocity. The amplitude of the small scale motion showed further increase after $\sim$ 2:30UT, and finally the filament erupted around 3:30UT. 

In order to quantify the small scale motions in the filament prior to the eruption, we made the histograms of the LOS velocity.

\section{RESULTS}  \label{sec:result}

\begin{figure}[ht!]
\plotone{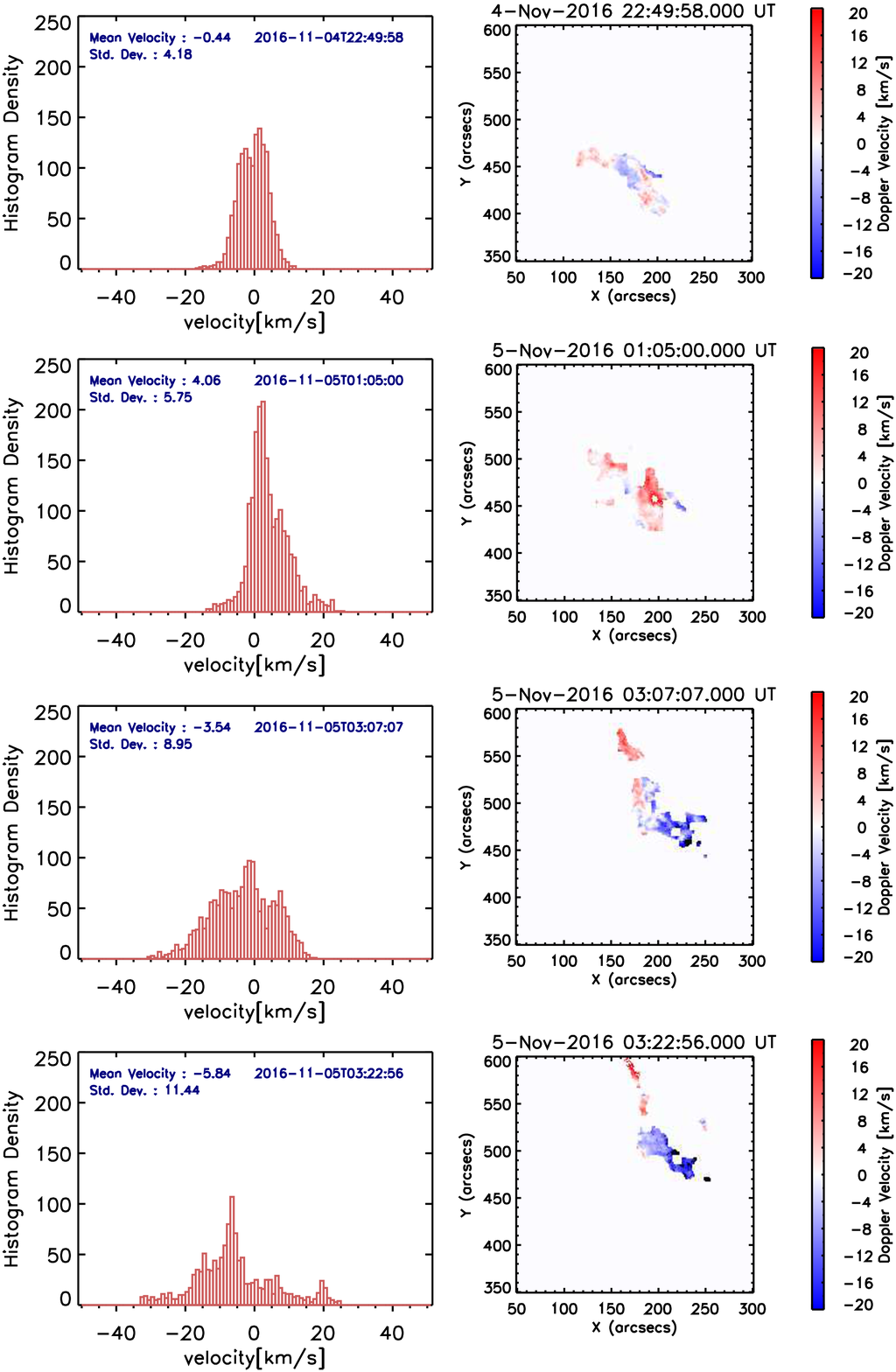}
\caption{\textit{Left} : The histograms of the LOS velocity images. Each histogram corresponds to the right image. The word ``Std. Dev.'' means standard deviation. \textit{Right} : 4 LOS velocity images inside the black squares of Fig.\ref{5types}. \label{4hist}}
\end{figure}

Figure \ref{4hist} shows the examples of the histograms and the corresponding LOS velocity images at 22:49UT on Nov. 4, 1:05UT, 3:07UT and 3:22UT on Nov. 5. 

The mean velocity and the standard deviation are also shown in the figure. One can recognize that the histograms are quite symmetric before the eruption, but the standard deviation increases with time and the histograms themselves become asymmetric.

\begin{figure}[ht!]
\gridline{\fig{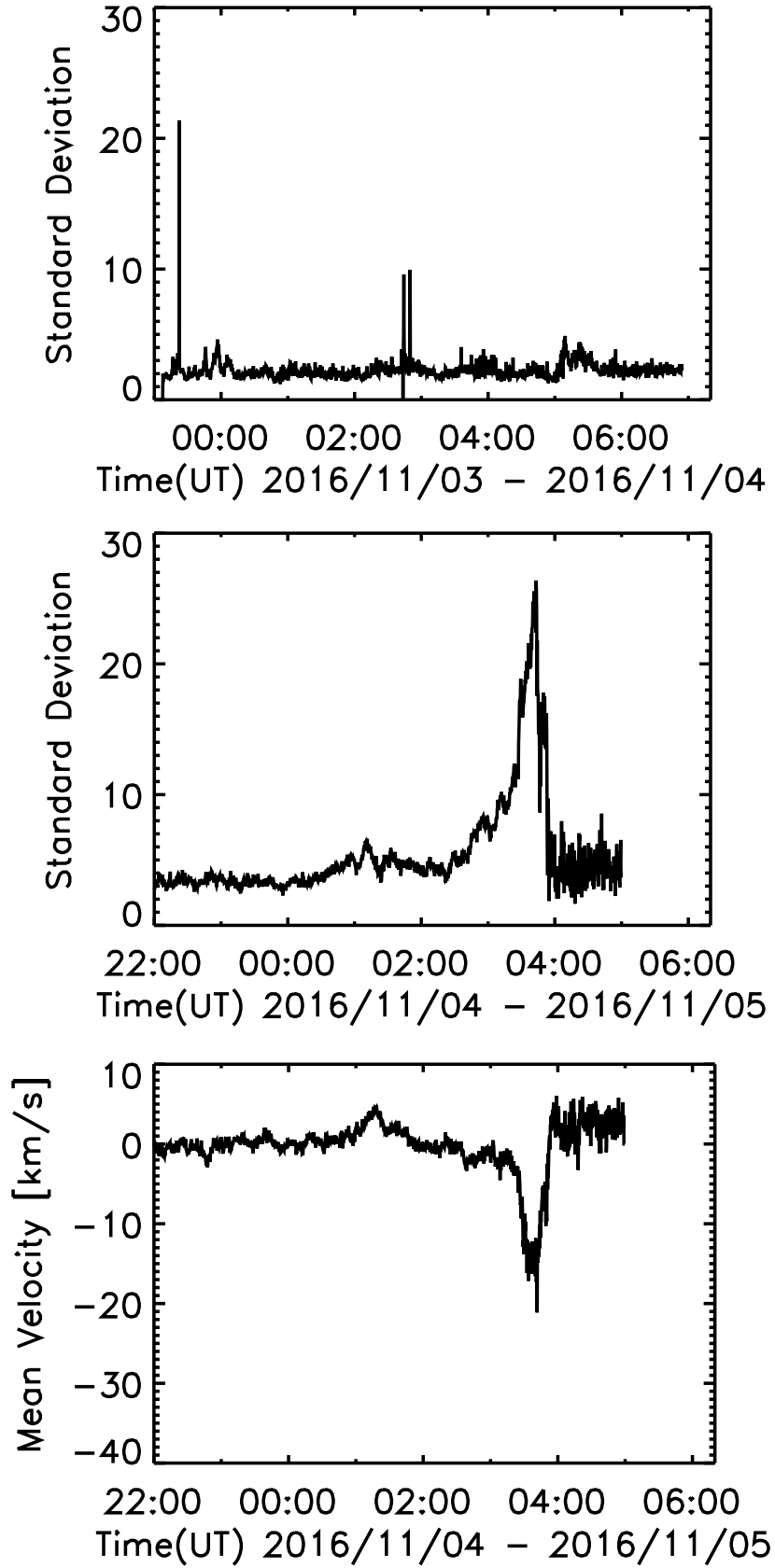}{0.6\textwidth}{}}
\caption{\textit{Top 2 panels} : The time transition of standard deviations of the histograms from 23:00 on November 3 to 7:00 on November 4 (above) and from 22:00UT on November 4 to 5:00UT on November 5 (below). \textit{Bottom panel} : The time transition of average of LOS velocities.\label{stdmean}}
\end{figure}

Figure \ref{stdmean} shows the temporal profile of the standard deviation of the histogram and the mean LOS velocity. 

For the standard deviation, the data from the previous day is also shown (from 23:00UT on November 3 to 7:00UT on November 4). The standard deviation stays almost constant around 2 $\sim$ 3 km s$^{-1}$ on the previous day (23:00UT on November 3 -- 7:00UT on November 4). At the beginning of the observation on the next day (22:00UT on November 4), the standard deviation slightly increases to 3 $\sim$ 4 km s$^{-1}$. It stays constant until $\sim$ 0:30UT and then gradually increase with a rate of 1.1 m s$^{-2}$ until it peaks at around 1:10UT. This peak in the standard deviation is associated with a positive peak of the mean velocity. Then the standard deviation starts to increase again at around 2:30UT with a rate of 2.8 m s$^{-2}$, and at around 3:10UT both the standard deviation increases sharply and also the mean velocity decreases sharply, leading to the onset of the eruption.

\section{Discussion} \label{sec:concl}

The SDDI/SMART allowed us to obtain the LOS velocity inside a filament with unprecedented detail.
The evolution of the standard deviation of the histogram for LOS velocity as the filament 
approaches the eruption provides an unique information on the physical state of the filament 
and may be regarded as a precursor of the eruption that can be used for the prediction of the eruption. 

The increase of the standard deviation with a rate of 2.8 m s$^{-2}$ during 2:30--3:10UT before the eruption may correspond to the slow-rise phase commonly observed before filament eruptions 
\citep{2004ApJ...602.1024S, 2006A&amp;A...449L..17I}. By investigating the H$\alpha$ center images in Fig. 3 one can recognize the global drift of the filament toward north-west in the plane of the sky. On the other hand, the weaker (1.1 m s$^{-2}$) increase in the standard deviation starting around 
00:00UT is not associated with a global drift of the filament. This may be regarded as the precursor of 
the onset of the slow-rise phase. 

It should be also noted that, the standard deviation during 22:00--00:00UT is almost constant, 
but its absolute value is slightly larger than that in the previous day. 
Here we present a possible interpretation of this. 
Small scale vertical motions have been commonly found in the high resolution observations 
of quiescent filaments \citep{2008ApJ...676L..89B, 2011Natur.472..197B}. 
Its physical origin is still uncertain, but one promising mechanism is 
the magnetic Rayleigh-Taylor instability \citep{2011ApJ...736L...1H, 2012ApJ...761..106H}. 
In the non-linear phase of the instability, the termination velocity of the rising plumes 
and sinking spikes is determined by the balance of the buoyancy (gravity) force and 
the drag force from the ambient plasma. It is likely that as the filament evolves toward 
the eruption, it expands and the magnetic field becomes weaker. This results in the 
decrease of magnetic drag on the plumes and spikes and thus increase in the average 
small velocity inside the filament. This hypothesis may be verified by statistical 
analyses of the quiescent filaments that shows the vertical motions. 

From the space weather point of view, filament eruptions can cause geomagnetic storms and bring us societal and economical impacts. It should be noted that the eruptions of quiescent filaments far from active regions,
which cannot be detected as a flare by in soft X-ray, may produce significant geomagnetic storm \citep{1996JGR...10113497M}. 
Therefore, the prediction of filament eruptions is of significant importance for mitigating the space weather hazard.

%The increase in the small scale velocity found in the present study may be regarded as a precursor of the filament eruptions.
%The H$\alpha$ and its line-wing observation like that by the SDDI is the promising way to monitor this precursor. 
%We emphasize that it is very important that this observation can be done from the ground. 
%Although the space observation has enormous advantages, the observatories in the space are vulnerable to space weather hazards themselves. Hence it is essential to keep a space weather monitoring and predicting system based on the data from the ground-based telescopes. 

The increase in the small scale velocity found in the present study may be regarded as a precursor of the lament eruptions.
Combining it with the other diagnostics for eruption such as the height of the filaments \citep{2001JGR...10625177F, 2007ApJ...668..533N}, period of oscillatory motion \citep{2006A&amp;A...449L..17I, 2007SoPh..246...89I, 2009ApJ...700.1658F} ,and the decay index of the coronal magnetic field \citep{2006PhRvL..96y5002K, 2015SoPh..290.1703M} will improve the prediction capability of solar eruptions. 
We emphasize that it is very important that the H$\alpha$ observation can be done from the ground-based telescopes.

%As for the eruption mechanism,  2 candidates can be thought according to the 304\AA images observed  by the Atmospheric Imaging Assembly on the Solar Dynamics Observatory, a flare has occurred near the active region NOAA 12605 around 12:40UT on November 4. The filament seems to be pressed after this flare 

\section*{Acknowledgement}
We thank A. Hillier for his comment on the magnetic Rayleigh-Taylor
instability in the filament. 
And we thank Prof. Shibata, Prof. Ichimoto, and Prof. Asai for the fruitful and meaningful discussions.
This work was partly supported by JSPS
KAKENHI Grant Numbers JP15H05814 and JP16H03955. We are grateful to the staff of Hida
Observatory for supporting the instrument development and daily observations. 
This work was also partly supported by the "UCHUGAKU" project of the Unit of Synergetic Studies for Space, Kyoto University.

%\bibliography{2017_Seki_et_al}
\end{document}